\documentclass[aip,jcp,reprint]{revtex4-1}
\usepackage{amsmath,amssymb,amsfonts}
\usepackage{graphics,subfigure}
\usepackage{epsfig}
\usepackage{color,soul}
\usepackage{graphicx}

\newcommand{\meanmu}{\bar{\mu}}
\newcommand{\mua}{\mu_a}
\newcommand{\mub}{\mu_b}
\newcommand{\workrat}{work ratio }
\newcommand{\eff}{efficiency }
\renewcommand{\eqref}[1]{Eq.~(\ref{#1})}
\newcommand{\een}{\textrm{I}}
\newcommand{\twee}{\textrm{II}}
\newcommand{\drie}{\textrm{III}}
\newcommand{\vier}{\textrm{IV}}
\newcommand{\p}{\mathcal{P}}
\newcommand{\m}{\bar{\mu}}
\newcommand{\maxp}{\textrm{mp}}

\begin{document}
\title{Efficiency at maximum power of a chemical engine}
\author{Hans Hooyberghs}
\email{hans.hooyberghs@fys.kuleuven.be}
\affiliation{Instituut voor Theoretische Fysica, KU Leuven, Celestijnenlaan 200D,  B-3001 Leuven, Belgium}
\author{Bart Cleuren}
\affiliation{Hasselt University, B-3590 Diepenbeek, Belgium}
\author{Alberto Salazar}
\affiliation{Instituut voor Theoretische Fysica, KU Leuven, Celestijnenlaan 200D,  B-3001 Leuven, Belgium}
\author{Joseph O. Indekeu}
\affiliation{Instituut voor Theoretische Fysica, KU Leuven, Celestijnenlaan 200D,  B-3001 Leuven, Belgium}
\author{Christian Van den Broeck}
\affiliation{Hasselt University, B-3590 Diepenbeek, Belgium}

\begin{abstract}
A cyclically operating chemical engine is considered that converts chemical energy into mechanical work. The working fluid is a gas of finite-sized spherical particles interacting through elastic hard collisions. For a generic transport law for particle uptake and release, the efficiency at maximum power $\eta_{\maxp}$ takes the form $1/2+c \,\Delta \mu + {\cal O}(\Delta \mu^2)$, with $1/2$ a {\em universal} constant and $\Delta \mu$ the chemical potential difference between the particle reservoirs. The linear coefficient $c$ is zero for engines featuring a so-called left/right symmetry or particle fluxes that are antisymmetric in the applied chemical potential difference. Remarkably, the leading constant in $\eta_{\maxp}$ is non-universal with respect to an exceptional modification of the transport law. For a nonlinear transport model we obtain $\eta_{\maxp}= 1/(\theta +1)$, with $\theta >0$ the power of $\Delta \mu$ in the transport equation. 
\end{abstract}
\maketitle
\section{Introduction}

Our starting point is the consideration of classic reversible cycles, such as that pertaining to the Carnot engine or chemical engine, and their finite-time realizations which yield a finite power output. For the (thermal) Carnot cycle this realization was studied by Curzon and Ahlborn \cite{Curzon1975}, while for an isothermal cycle, converting chemical work to mechanical work, a particular finite-time realization was examined more recently by Chen {\em et al.} \cite{Chen}. In both cases the efficiency at maximum power, relative to the ideal efficiency, was shown to be precisely 1/2, in the limit of a small temperature difference (for the thermal cycle) or a small chemical potential difference (for the isothermal chemical cycle) between the two reservoirs that serve as heat and/or particle source and sink for the working fluid. This result was subsequently derived based upon general arguments for the thermal engine \cite{ChrisVdB,devos,esposito2009,esposito2010}. Moreover, the issue of efficiency at maximum power has since then been revisited in a number of papers for different thermodynamic cycles and models \cite{udo2008,udo2011,udo2012,esposito2012,chris2012,imperato2012}.

In this work we complement these previous studies. Besides considering the finite-time realizations of the processes, necessary to obtain a non-zero power output, we also consider the {\em finite-size versions of the particles} constituting the working fluid, instead of the usual point-like ideal gas particles of zero size. This refinement is interesting for the chemical cycle, in which the transport coefficient involves the effusion and diffusion of the particles that make up the working fluid. In contrast, for the thermal cycle with heat exchange and particle conservation it is not relevant, since the transport coefficient in that case is the thermal conductivity of the engine wall, which is a solid-state property independent of the nature of the working fluid. The case of the thermal cycle is discussed in Appendix A.

The objectives of this work are twofold. Firstly, it introduces a concrete implementation of a chemical engine that cyclically converts chemical energy into mechanical work. During one cycle, steps involving isothermal uptake/release of particles by the working fluid are alternated by isothermal expansion/compression steps. Secondly, such a concrete model allows us to obtain explicit expressions for the transport coefficients used in the earlier derivations of the efficiency at maximal power of an isothermal chemical engine, as described, e.g., in \cite{Chen}. It becomes possible now to check whether or not those derivations are self-consistent, in the following sense. For calculating the efficiency at maximum power, the formal expressions for the power were extremalized with respect to chemical potential differences, while keeping the transport coefficients constant without explicit justification. In contrast, our present approach allows one to assess the validity of the classic calculations by inspecting whether the featured transport coefficients are indeed independent of the variables with respect to which differentiation is performed. Finally, we mention that our chemical engine serves as a paradigm of the thermodynamics of chemical to mechanical energy conversion.

\section{The chemical cycle}
\begin{center}
\begin{figure}[t]
\includegraphics[width = .4\textwidth]{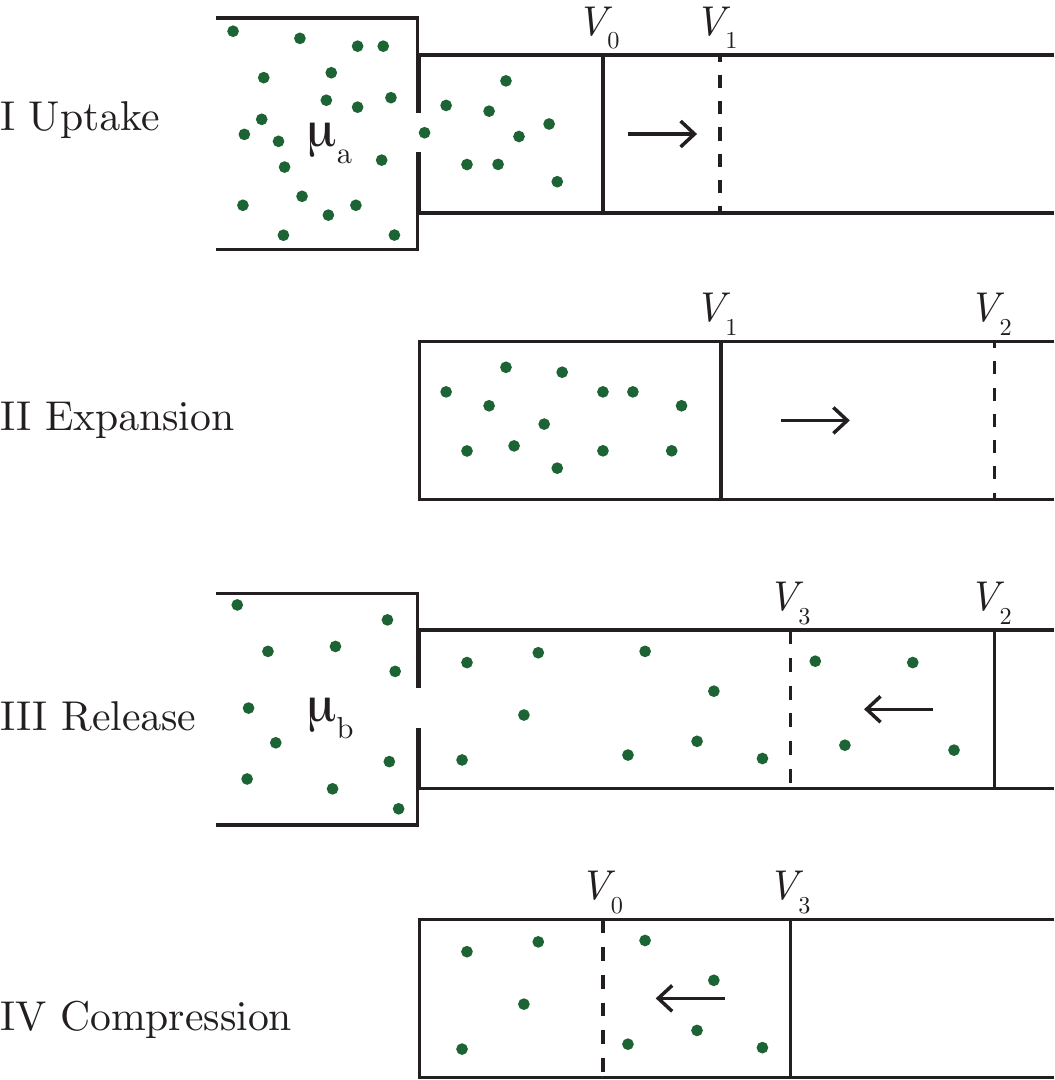}\caption{Sketch of the four different phases during the reversible chemical cycle: I) isothermal uptake of particles, II) isothermal and isocardinal ($N$ = constant) expansion, III) isothermal release of particles and IV) isothermal and isocardinal compression. Details are given in the text.\label{figuurCycle}}
\end{figure}
\end{center}

During the chemical cycle an auxiliary system is brought in contact with different particle and/or thermal reservoirs. The cycle has four different phases as shown in Figure~\ref{figuurCycle}. The auxiliary system is a cylinder which is sealed on one end by a fixed wall and by a movable piston on the other end. The fixed wall has a valve that can be opened or closed, depending on the phase of the cycle.

The working fluid consists of (mono-atomic) spherical particles of finite diameter $\sigma$, which interact solely via hard-sphere collisions. The Clausius equation of state (EOS) for such a fluid (i.e., the ``mean-field" or van der Waals equation of state without inter-particle attractions \cite{vanderWaals}) is
\cite{Tabor,Callen}
\begin{equation}
 p (V-Nv_0) = N k_BT\label{eos1},
\end{equation}
where $p$ is the pressure, $N$ the number of particles, $k_B$ the Boltzmann constant and  $T$ the temperature. The so-called free volume $V- Nv_0$ is the physical volume minus the co-volume or excluded volume. Each particle excludes a volume $v_{ex} = 4\pi\sigma^3/3$ to the centers of other particles, and hence $N v_0 \equiv N v_{ex}/2$ is a low-density (or one-dimensional \cite{Tonks}) approximation to the total excluded volume for the hard-sphere fluid. Within this approximation the hard-sphere fluid is henceforth in short referred to as a {\em Clausius gas}. 
We can rewrite the Clausius EOS in the form 
\begin{equation}
p = \frac{\rho}{1-\rho v_0}k_BT,\label{eos2}
\end{equation}
with $\rho$ the number density. 

The Clausius EOS is exact for hard rods in one dimension (for which $v_0 = \sigma$), and can serve as a mean-field like approximation for hard particles (disks or spheres) in higher dimensions \cite{Tonks}. The chemical potential of the Clausius gas contains an additional term relative to that of an ideal gas, due to the finite volume of the particles \cite{Tabor,Callen,Tonks}:
\begin{equation}
\mu = k_B T\left\{\log\left(\frac{\rho\Lambda^3}{1-\rho v_0}\right)+\frac{\rho v_0}{1-\rho v_0}\right\},\label{mu}
\end{equation}
with $\Lambda$ the thermal de Broglie wavelength and where we assumed a three-dimensional space.
\begin{center}
\begin{figure}[t]
\includegraphics[width = .4\textwidth]{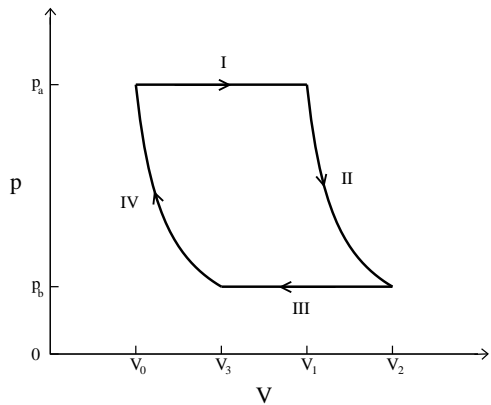}\caption{$p-V$ diagram for the reversible chemical cycle. \textcolor{black}{The temperature $T$ is constant throughout.} Path I is an isothermal and isobaric expansion, with particle uptake. Path II is an isothermal and isocardinal ($N$ = constant) expansion. Path III is an isothermal and isobaric compression, with particle release. Path IV is an isothermal and isocardinal compression. The enclosed area is the total mechanical work done by the engine. It equals the total chemical work $(\mu_a-\mu_b) \Delta N$ done on the engine by transferring $\Delta N$ particles from a high chemical potential reservoir to one with a lower chemical potential. \label{figuurpV}}
\end{figure}
\end{center}
\begin{table*}
\caption{Overview of the thermodynamic quantities in the reversible cycle. For the finite-time (irreversible) cycle one may obtain the various energy exchanges simply by considering a reversible cycle operating between particle reservoirs at modified chemical potentials $\mu_a^*$ and $\mu_b^*$, and modified densities $\rho_a^*$ and $\rho_b^*$, respectively. Note that in this point of view the chemical energy wasted ``outside" the engine, in transferring $\Delta N$ particles from  $\mu_a$ to $\mu_a^*$ and from $\mu_b^*$ to $\mu_b$ must be taken into account when considering the efficiency.\label{tabelrev}}
\begin{tabular}{c|c|c|c|c}
Phase & $\Delta U_j$  & $W_{j}$&$W_{j}^{chem}$ & $Q_j$\\
\hline
I& $\frac{3}{2}\Delta N k_B T$ &$-\frac{k_B T}{1-\rho_a v_0}\Delta N$&$\mua \Delta N$& $\Delta N\left(\frac{3}{2}k_BT-\mu_a+\frac{kT}{1-\rho_a v_0} \right)$\\
II&0&$-(N+\Delta N)\left( (\mua-\mub) + v_o k_B T \left(\frac{-\rho_a}{1-\rho_av_0} +\frac{\rho_b}{1-\rho_bv_0}\right)\right)$&0&$ -W_{\twee}$ \\
III&$-\frac{3}{2}\Delta N k_B T$&$\frac{k_B T}{1-\rho_b v_0}\Delta N$&$-\mub \Delta N$&$\Delta N\left(-\frac{3}{2}k_BT+\mub-\frac{kT}{1-\rho_b v_0} \right)$\\
IV&0&$N\left( (\mua-\mub) + v_o k_B T \left(\frac{-\rho_a}{1-\rho_av_0} +\frac{\rho_b}{1-\rho_bv_0}\right)\right)$&0&$-W_{\vier}$\\
\hline
Total	&$0$			&$-\Delta N(\mu_a-\mu_b)$&$\Delta N(\mu_a-\mu_b)$	&$0$\\\hline
\end{tabular}
\end{table*}

\subsection{Reversible cycle}
The four phases of the reversible cycle are now described in detail. During each phase $j\in \{I,II,III,IV\}$ of the cycle, the change of energy $\Delta U_j$ of the auxiliary system has three contributions: 
\begin{equation}\label{law1}
\Delta U_j=Q_j+ W_j+\mu_j \Delta N_j
\end{equation}
with $Q_j$ the heat absorbed by the gas, $W_j$ the mechanical work done on the gas due to the piston motion, and $W_j^{chem}=\mu_j \Delta N_j$ the chemical work input from the reservoir. \\
$\bullet$ Phase I: Particle uptake\\
Initially, the cylinder containing the working gas has a volume $V_0$ in which $N$ particles are present. \textcolor{black}{Initially, $N = \rho_a V_0$.} Subsequently, the valve to the first particle reservoir ($\mu = \mu_a$ and $\rho = \rho_a$) is opened and $\Delta N$ particles diffuse (or effuse) into the cylinder. The piston moves outwards and a final volume $V_1$ is reached. Throughout the process, the density in the cylinder remains constant at $\rho_a$. Also the temperature is kept constant by virtue of a heat input from a thermal reservoir. Consequently, the pressure remains constant at $p_a = \frac{\rho_a}{1-\rho_a v_0}k_BT$.  The mechanical work done on the gas is, with $\Delta N = \rho_a(V_1 - V_0)$, 
\begin{equation}
W_{\een} = -\frac{k_BT}{1-\rho_a v_0}\Delta N,
\end{equation}
while the chemical work done on the gas is 
\begin{equation}
W_{\een}^{chem} = \mu_a \Delta N.
\end{equation}
$\bullet$ Phase II: Isothermal expansion\\
During the second phase, the valve to the reservoir is closed. The gas undergoes an isothermal expansion from volume $V_1$ to $V_2$, while the density decreases from $\rho_a$ to $\rho_b$. The systems remains at constant temperature owing to a net heat input. The mechanical work done on the gas is
\begin{multline}
W_{\twee}=-(N+\Delta N)(\mua-\mub) \\-(N+\Delta N)v_o k_B T \left(\frac{-\rho_a}{1-\rho_av_0} +\frac{\rho_b}{1-\rho_bv_0}\right).
\end{multline}
Since the number of particles is fixed in this process, for which we coin the succinct term ``isocardinal"\cite{isocardinal}, the chemical work equals zero.\newline
$\bullet$ Phase III: Particle release\\
The valve is opened to the second reservoir ($\mu = \mu_b < \mu_a$ and $\rho=\rho_b$)  at the start of the third phase and $\Delta N$ particles diffuse from the working gas to the reservoir. The piston moves inwards until the volume $V_3$ is reached. The temperature, density $\rho_b$ and pressure $p_b$ are constant throughout the particle release. The mechanical work performed on the gas by the piston and the chemical work done on the gas are
\begin{eqnarray}
W_{\drie} &=& \frac{k_BT}{1-\rho_b v_0}\Delta N,\\
W^{chem}_{\drie} &=&-\mu_b \Delta N.
\end{eqnarray}
$\bullet$ Phase IV: Isothermal compression\\
Finally, the gas undergoes an isothermal and isocardinal compression from $V_3$ to $V_0$. The valve is closed and the piston moves inwards to regain the initial state with volume $V_0$ and density $\rho_a$. The chemical work is zero, while the mechanical work performed on the gas is 
\begin{multline}
W_{\vier}=N(\mua-\mub) \\ + N v_o k_B T \left(\frac{-\rho_a}{1-\rho_av_0} +\frac{\rho_b}{1-\rho_bv_0}\right).
\end{multline}
\noindent
The $p$-$V$ diagram of the process is shown in Figure \ref{figuurpV}. An overview of the thermodynamic quantities in the cycle is presented in Table \ref{tabelrev}. The heat exchange $\Delta Q$ is obtained using the first law of thermodynamics \cite{measurable}.

The \workrat of the chemical cycle, called ``efficiency'' \cite{Chen} (but for alternative definitions, see \cite{shibata}),  is defined as
\begin{equation}
\eta = \frac{-\sum W_j}{\sum W_j^{chem}},
\end{equation}
where the sum is over all 4 phases.
Because the engine is isothermal and reversible, $T\Delta S_{tot} = \sum Q_{j} = 0$, where the subscript $tot$ refers to the total over all 4 phases. The data in Table \ref{tabelrev} are in accord with this property. An application of the first law of thermodynamics then straightforwardly yields that the \workrat equals 1. Since these arguments are independent of the details of the engine, the \workrat equals 1 for all reversible isothermal cycles. For the sake of conformity with familiar nomenclature, we will henceforth also adopt the term ``efficiency" for $\eta$. 

\subsection{Irreversible cycle: efficiency at maximum power}
During the particle uptake and release phases of the reversible cycle, the chemical potential of the working fluid is at all times equal to the chemical potential of the respective reservoirs. This implies that the particle transport is infinitely slow. In order to speed up the cycle, a drop in chemical potential (or density) between the reservoir and working fluid must be introduced. We make the following assumptions. We assume that the temperature is kept constant throughout. In order to allow the particles to leave the first reservoir and to enter the working gas, we assume a drop in chemical potential from $\mua$ to $\mua^*$. The intake of particles takes place at constant chemical potential $\mua^*$ and therefore at constant density $\rho_a^*$, compatible with the temperature $T$ and the value of the chemical potential $\mu = \mua^*$, according to \eqref{mu}. Similarly, the release of particles in the third step of the cycle is assumed to take place at constant chemical potential $\mub^*$ and constant density $\rho_b^*$, again related through \eqref{mu}. In order to permit the flow of particles from the working gas into the second reservoir, we assume a chemical potential drop from $\mub^*$ to $\mub$. Finally, without loss of generality, the number of particles transferred during the uptake and release phases remains equal to $\Delta N$. And so the volumes are adapted during the different phases, e.g., $\rho_a^*(V_1^*-V_0)=\Delta N$. We note that both chemical potential drops can be considered to occur outside the cycle. This is consistent with the application of the first law of thermodynamics to each phase of the cycle. In particular, the thermodynamic quantities given in Table \ref{tabelrev} are {\em all} to be replaced by their counterparts in the irreversible cycle (that is, $\mu_i \rightarrow \mu_i^*$ and $\rho_i\rightarrow\rho_i^*$, with $i = a$ or $b$).

The \eff is now
\begin{equation}\label{effratio}
\eta = \frac{\mua^*-\mub^*}{\mua - \mub},
\end{equation}
in accord with the result found in \cite{Chen}. We have $\eta <1$, since only a fraction of the total available chemical work is done on the gas and converted to mechanical work done by the gas. More precisely, the amounts $(\mua-\mua^*)\Delta N$ and $(\mub^*-\mub)\Delta N$ of chemical work are ``wasted'' during uptake and release of the particles.

The power output of the cycle is given by
\begin{equation}
P = \frac{-\sum W_j}{\tau},
\end{equation}
where $\tau$ is the (finite) duration of the cycle. We now derive the efficiency at maximal power, in a way similar to the calculational scheme of Curzon and Ahlborn  for the thermal cycle \cite{Curzon1975} and that of Chen {\em et al.} \cite{Chen} for the chemical cycle. The dynamics of the cycle is not yet determined. When the two reservoirs are close to equilibrium, i.e., \textcolor{black}{when $(0 <)\; \mu_a-\mu_b \ll \;k_BT$,} the particle flux is well described by \cite{Callen}
\begin{equation}\label{diffusion}
\frac{dN}{dt}=\lambda_i(\mu_i-\mu_i^*),
\end{equation}
where the $\lambda_i \;(>0)$ are essentially temperature-dependent transport coefficients. As we show in the next section, these coefficients do not depend, or depend only weakly, on the chemical potentials. Note that $\frac{dN}{dt} >0 $ $ (<0)$ during particle uptake (release).

The total time to complete phases I (particle uptake) and III (particle release)  is then 
\begin{equation}
\tau_1+\tau_3=\frac{\Delta N}{\lambda_a(\mua-\mua^*)}+\frac{\Delta N}{\lambda_b(\mub^*-\mub)}\label{tau}.
 \end{equation}
In analogy with the Curzon-Ahlborn approach, the time required for the isocardinal phases II and IV is set to $(q-1)(\tau_1+\tau_3)$, where $q $ is a constant. Consequently, the total time to complete a full cycle is simply $q (\tau_1+\tau_3)$. The power output of the cycle is 
\begin{equation}
P = \frac{1}{q}\frac{\mua^*-\mub^*}{\frac{1}{\lambda_a(\mua-\mua^*)}+\frac{1}{\lambda_b(\mub^*-\mub)}}.
\end{equation}
Given the values for $\mu_a$ and $\mu_b$, we now maximise the power with respect to $\mu_a^*$ and $\mu_b^*$. This can be done analytically and yields:
\begin{equation}
\mu_a^{*}=\m+\frac{\alpha(\mu_a-\mu_b)}{2(1+\alpha)} \;\;\; ; \;\;\; \mu_b^{*}=\m-\frac{\mu_a-\mu_b}{2(1+\alpha)}
\end{equation}
where we introduced the average chemical potential in the reservoirs:
\begin{equation}
\m = \frac{\mu_a+\mu_b}{2}
\end{equation}
and $\alpha^2=\lambda_a/\lambda_b$. Note 
\begin{equation}
\Delta \mu^* \equiv \mu_a^{*} - \mu_b^* = \frac{\mu_a-\mu_b}{2}\equiv \frac{\Delta \mu}{2 }.
\end{equation}
The resulting efficiency at maximum power is found to be
\begin{equation}
\eta_{\maxp} = \frac{1}{2},\label{etaexp}
\end{equation}
which is exact within linear response theory.

Since the linear transport law \eqref{diffusion} is strictly speaking only valid when the differences in chemical potential are small, the question arises  how robust this number $1/2$ is against modifications of the transport law. The modifications we have in mind obviously include taking into account non-linearity and allowing for the presence of higher-order terms $(\mu_i-\mu_i^*)^n$, with $n$ odd, but we also think of situations in which the thermodynamic variables are not the chemical potentials themselves but smooth functions of them. For example, when the linear dimension of the opening between working fluid and reservoir is smaller than the mean free path of the  particles in the gas, transport occurs through effusion. The (linear) particle current is then naturally expressed as being proportional to the difference in densities \cite{effusion}:
\begin{equation}\label{effu}
\frac{dN}{dt}=A\sqrt{\frac{k_BT}{2\pi m}}\left(\rho_i-\rho_i^*\right),
\end{equation}
with $A$ the area of the opening and $m$ the mass of a single particle. Further, if the working fluid and the reservoir are connected by a valve of opening cross-section $A$ and finite length $L$, over which the density varies smoothly, the (linear) particle transport is limited by diffusion and described by Fick's law,
\begin{equation}
\frac{dN}{dt} = - DA\frac{d\rho}{dx},
\end{equation}
with $D$ the diffusion coefficient of the fluid, which depends on molecular constants, number density and temperature. This expression involving the density gradient can be rewritten in terms of the chemical potential gradient, using \eqref{mu}, 
\begin{equation}
\frac{dN}{dt} = - \frac{D\rho}{k_BT} (1-\rho v_0)^2 A\frac{d\mu}{dx} \approx - \frac{D\rho}{k_BT}  A\frac{d\mu}{dx},
\end{equation}
where the final expression is a low-density approximation. 
Simple kinetic theory of gases implies that $D$ is proportional to the product of the thermal velocity $v_{\rm rms}$ and the mean free path $l_{\rm mf}$ of the molecules of mass $m$, so that, for our Clausius gas, 
\begin{equation}
D \propto l_{\rm mf} v_{\rm rms}\propto \frac{1}{\rho \sigma^2}\sqrt{\frac{k_BT}{m}},
\end{equation}
Consequently, the product $D \rho$ depends mainly on molecular constants and temperature, and only weakly on density. We conclude that, in the low-density approximation, the transport coefficient $\lambda$ in \eqref{diffusion} is given by
\begin{equation}
\lambda  \propto \frac{A}{\sigma^2 L}\frac{1}{\sqrt{mk_BT}},
\end{equation}
with a proportionality constant of order unity.
We observe that, for a dilute gas, $\lambda$ depends only on temperature, molecular constants and the dimensions of the valve. Consequently, for a dilute gas, it is justified to keep $\lambda$ constant when considering variations of the chemical potential at constant temperature. For a left/right symmetric cycle (with input and output valves of equal size), $
\lambda_a =   \lambda_b$.

Note that when the particle size $\sigma$ is taken to zero, the point particles do not hinder each other during entry or escape and only effusion limits the current. Whether or not diffusion can be neglected depends on the ratio of the volume of an ``escape tube", $\sigma^2 L$, to the available volume per particle $1/\rho$. If the former is much smaller than the latter, the tube is ``free" and the particle exchange between engine and reservoir is well described by effusion alone. Conversely, transport is diffusion-limited when traffic in the tube is dense ($ \rho \sigma^2 L \gg 1$). We can capture the crossover between these two limits by writing down an effective transport coefficient
\begin{equation}
\lambda_{\rm eff}  \approx \frac{A}{\sigma^2 L + 1/\rho}\frac{1}{\sqrt{mk_BT}},
\end{equation}
where we neglected constants of order unity in the expression.

In order to encompass all such situations, and to retain full generality in the subsequent discussion, we propose the following generic transport law:
\begin{equation}\label{flux}
\frac{dN}{dt}=\kappa_{i}f(\mu_{i},\mu_{i}^{*}) \;\;\;\;(i\in\{a,b\}).
\end{equation}
The (effective) transport coefficients are now by definition independent of the chemical potential \cite{valve} and the function $f$ is required to satisfy two conditions: i) skew-symmetry, $f(x,y)=-f(y,x)$, and  ii) both first-order partial derivatives do not vanish at $x=y=\meanmu$. Note that \eqref{effu} provides an example if the form of $f(x,y) = F(x)-F(y)$. Also note that the second condition rules out certain purely nonlinear transport laws, which are interesting for us and therefore merit a separate discussion (see further). 

The maximization of the power can no longer be done exactly, and we resort to a series expansion. Since $\mu_{a} \geq \mu_{a}^{*}\geq \mu_{b}^{*} \geq \mu_{b}$ we  compose the following small parameter
\begin{equation}
\varepsilon=\frac{\mu_{a}-\mu_{b}}{\mu_{a}+\mu_{b}}.
\end{equation}
and define:
\begin{equation}\label{eq:subs}
\begin{array}{lcl}
\mu_{a}=\m(1+\varepsilon) &;&\mu_{b}=\m(1-\varepsilon)\\
\mu_{a}^{*}=\m(1+x(\varepsilon))&;&\mu_{b}^{*}=\m (1+y(\varepsilon)).
\end{array}
\end{equation}
The efficiency can then be expressed as:
\begin{equation}\label{eta}
\eta = \frac{x(\varepsilon)-y(\varepsilon)}{2\varepsilon}.
\end{equation}
To proceed, we calculate $\partial P/\partial x$ and $\partial P/\partial y$ and then substitute the series expansions $x(\varepsilon)=a_1 \varepsilon+a_2 \varepsilon^2\ldots$ and $y(\varepsilon)=b_1 \varepsilon+b_2 \varepsilon^2\ldots$. The coefficients are determined by the extremality conditions $\partial P/\partial x =0$ and $\partial P/\partial y =0$. This leads to the following results:
\begin{equation}
x(\varepsilon)=\frac{\alpha \varepsilon }{1+\alpha}
+\frac{(1+3\alpha)f^{2,0}}{4(\alpha+1)^2 f^{1,0}}\m \varepsilon^2+\ldots
\end{equation}
and
\begin{equation}
y(\varepsilon)=-\frac{\varepsilon }{1+\alpha}
+\frac{\alpha  (\alpha +3)f^{2,0}}{4 (\alpha +1)^2
   f^{1,0}}\m \varepsilon^2+\ldots
\end{equation}
with $f^{i,j}\equiv \partial_x^{i}\partial_y^{j}f(x,y)\vert_{(\m,\m)}$ and $\alpha^{2}=\kappa_a/\kappa_b$ is a measure of the asymmetry between the flux constants $\kappa_a$ and $\kappa_b$.
Substituting these results into the efficiency and again performing a series expansion in $\varepsilon$ leads to
\begin{widetext}
\begin{eqnarray}\label{effMP}
\eta_{\maxp}=\frac{1}{2}+\frac{(1-\alpha)f^{2,0}}{8(1+\alpha)f^{1,0}}\m\varepsilon
-\frac{
3(1+4\alpha+\alpha^2)(f^{2,0})^2+6(\alpha^2+1)f^{1,0}f^{1,2}-2(1+\alpha)^2 f^{1,0}f^{3,0}}{48(1+\alpha)^2(f^{1,0})^{2}}(\m\varepsilon)^{2} + {\cal O}((\m\varepsilon)^{3}).
\end{eqnarray}
\end{widetext}
Note that $\m\varepsilon=\Delta\mu/2$. 

Interestingly, there are two circumstances under which the first-order correction to the universal constant vanishes. Obviously, when the flux constants are equal, $\kappa_a=\kappa_b$ and hence $\alpha =1$, the first-order term vanishes. Furthermore, all odd-order terms vanish. The intuitive explanation goes as follows. When $\kappa_a=\kappa_b$ the system is spatially symmetric so that simultaneously running the cycle backward and switching $\mu_a \leftrightarrow \mu_b$ must lead to the same efficiency at maximum power. This is called left/right symmetry. Next, even when the engine is {\em not} left/right symmetric, the first-order correction vanishes when $f(x,y)$ depends only on the difference $x-y$, which covers a wide class of systems. In this case $f^{2,0} = 0$ because for an antisymmetric function all even derivatives vanish at the origin. We verified analytically that, as a consequence hereof, all odd-order corrections vanish. The efficiency at maximum power now is, with $f(x,y)=\tilde f(x-y)$,
\begin{eqnarray}\label{effMPsim}
\eta_{\maxp}=\frac{1}{2}-\frac{
(\alpha^2-\alpha +1) \tilde f '''(0)}{12(1+\alpha)^2\tilde f'(0)}(\m\varepsilon)^{2} + {\cal O}((\m\varepsilon)^{4}).
\end{eqnarray}
If, in addition, the system is left/right symmetric, we obtain
\begin{eqnarray}\label{effMPsimsim}
\eta_{\maxp}=\frac{1}{2}-\frac{
\tilde f '''(0)}{48\tilde f'(0)}(\m\varepsilon)^{2} + {\cal O}((\m\varepsilon)^{4}).
\end{eqnarray}

\subsection{Purely nonlinear transport laws}
In this section we test the robustness of \eqref{effMP} with respect to an exceptional modification of the transport equation. As a first example, we assume that there can exist circumstances under which the linear transport coefficient $\lambda$ (cf. \eqref{diffusion}) vanishes, while higher-order terms survive. For concreteness, we assume the following transport {\em model}
\begin{equation}\label{NLdiffusion}
\frac{dN}{dt}=\omega_i(\mu_i-\mu_i^*)^n,\,\,\,\mbox{with} \,\, n = 3, 5, ... 
\end{equation}
The extremality conditions are again analytically soluble, and yield:
\begin{eqnarray}
\mu_a^*&=&\mu_a-\frac{n(\mu_a-\mu_b)}{(1+n)(1+\gamma)}\;\;;\\
\mu_b^*&=&\mu_b+\frac{n(\mu_a-\mu_b)\gamma}{(1+n)(1+\gamma)}
\end{eqnarray}
with $\gamma=(\omega_a/\omega_b)^{1/(1+n)}$. Note 
\begin{equation}
\Delta \mu^*  =  \frac{\Delta \mu}{1+n}.
\end{equation}
Substitution in \eqref{effratio} leads to
\begin{equation}
\eta_{\maxp} = \frac{1}{n+1},
\end{equation}
which is actually valid for all odd values of $n\, ( \,\geq1)$.
Specifically, for a purely cubic nonlinearity ($n=3$), we obtain 
\begin{equation}
\eta_{\maxp} = \frac{1}{4}.
\end{equation}
Hence the efficiency at maximal power is non-universal with respect to this exceptional modification of the transport law and decreases with increasing order of nonlinearity.

As a second example, we consider a generalization of the previous case and postulate the transport equation
\begin{equation}\label{NL2diffusion}
\frac{dN}{dt}=\omega_i \;\mbox{sgn}(\mu_i-\mu_i^*) \mid \mu_i-\mu_i^* \mid^{\theta},\,\,\,\mbox{with} \,\, \theta > 0,
\end{equation}
with sgn the sign function, which provides the antisymmetry in the variable $\mu_i-\mu_i^*$. Note that this model includes {\em sublinear} transport laws ($\theta <1$). In general the transport equation is singular (except for integer and odd $\theta$, which is the case considered in the first example). Also for this more general model the efficiency at maximum power can be calculated exactly, following the foregoing steps without noteworthy modifications. The result is
\begin{equation}
\eta_{\maxp} = \frac{1}{\theta+1}.
\end{equation}
For sublinear transport laws this implies $\eta_{\maxp} > 1/2$, and, remarkably, the maximum efficiency can tend to 1 when the transport law approaches a step function. One has to keep in mind, however, that to our knowledge it has not yet been shown that the modified transport laws proposed in this section are physically possible. 

\section{Weak Dissipation Limit}
We now analyse the previous finite-time cycle from the perspective of weak dissipation. This analysis is based on the framework introduced in \cite{cvdbWD}, and goes as follows. During each step $j$ of the cycle, the change of energy $\Delta U_j$ of the auxiliary system has three contributions (cf. \eqref{law1}), 
\begin{equation}
\Delta U_j=Q_j+ W_j+\mu_j \Delta N_j.
\end{equation}
The entropy change $\Delta S_j$ can be written as
\begin{equation}
\Delta S_j=\Delta_e S_j+\Delta_i S_j=\frac{Q_j}{T}+\Delta_i S_j,
\end{equation}
with $\Delta_e S_j$ the entropy exchange with the environment and $\Delta_i S_j \geq 0$ the internal, irreversible entropy production. Since the auxiliary system returns to its initial state after one cycle we have $\sum \Delta U_j=\sum \Delta S_j=0$. The output power and efficiency of the cycle are then, respectively,
\begin{equation}\label{eq:powerWD}
\p=\frac{-\sum W_{j}}{\sum \tau_j}=\frac{\sum\left(\mu_j \Delta N_j-T \Delta_i S_j\right)}{\sum \tau_j}
\end{equation}
and
\begin{equation}
\eta=\frac{-\sum W_{j}}{\sum W_{j}^{chem}}=1-\frac{T\sum \Delta_i S_j}{\sum \mu_j \Delta N_j}.
\end{equation}
In the general framework described in \cite{cvdbWD}, it is argued that close to the reversible limit, the dissipation associated with the irreversible entropy production is inversely proportional to the operation times,
\begin{equation}
T\Delta_i S_j =\frac{\sigma_j}{\tau_j}+{\cal O}(1/\tau_j^2).
\end{equation}
Maximising the power with respect to the operation times gives:
\begin{equation}
\tau_i=\frac{2\sqrt{\sigma_i}\sum \sqrt{\sigma_j}}{\sum \mu_j \Delta N_j}
\end{equation}
and yields an efficiency at maximum power $\eta_{mp}=1/2$, irrespective of further details of the cycle.\\
The work-to-work cycle presented above nicely fits into this framework. In order not to overburden the notation, we now set $v_0=0$. In the first phase of the cycle an amount $\Delta N (\mu_a - \mu_a^*)$ is not used to do mechanical work, and a similar accident happens in phase III. The corresponding internal irreversible entropy productions are
\begin{equation}
T\Delta_i S_I=(\mu_a-\mu_a^{*})\Delta N \;\; ; \;\; T\Delta_i S_{III}=(\mu_b^{*}-\mu_b)\Delta N
\end{equation}
and $T\Delta_i S_{II}=T\Delta_i S_{IV}=0$, since during those phases the entropy change is just the (reversible) heat exchange divided by $T$. \textcolor{black}{For the uptake phase, we need to express $\mu_a - \mu_a^*$ in terms of the duration $\tau_1$. Since the chemical potentials remain constant during each phase, we can simply integrate \eqref{flux}, which gives}
\begin{equation}
f(\mu_a,\mu_a^*)=\frac{\Delta N}{\kappa_a \tau_1}.
\end{equation}
This expression can be solved for $\mu_a^*$ to first order in $1/\tau_1$,
\begin{equation}
\mu_a^*=\mu_a-\frac{\Delta N}{\kappa_a \partial_x f(x,y)\vert_{(\mu_a,\mu_a)}\tau_1}+\ldots,
\end{equation}
where we made use of the skew symmetry of $f$.
Hence we find
\begin{equation}
T\Delta_i S_I\approx\frac{\Delta N^2}{\kappa_a \partial_x f(x,y)\vert_{(\mu_a,\mu_a)} \tau_1},
\end{equation}
which is precisely the weak dissipation condition. A similar expression is obtained for $T\Delta_i S_{III}$. So once again we recover the universal value $1/2$ for the efficiency at maximum power. Note that this derivation breaks down for the purely nonlinear transport model (Section II.C), since $\partial_x f(x,y)\vert_{(u,v)}$ vanishes identically for $u=v$.
\section{Conclusion}
In this paper we have achieved the following. \\
i) We have proposed a paradigm for the thermodynamics of energy conversion in the form of a concrete implementation of a chemical engine that cyclically converts chemical energy into mechanical work. \\
ii) We have complemented the finite-time (and consequently finite-power) analysis of a chemical cycle with a finite-size description of the particles constituting the working fluid. Our treatment goes beyond the usual point-particle description, with which one can only discuss the uptake and release of particles as an effusion process. We have thus provided a minimal setting for invoking transport coefficients, such as the diffusion constant, which -- of course -- exist by virtue of the fact that the particles have a finite diameter. Starting from the Clausius equation of state we have given explicit expressions for the thermodynamic quantities characterizing the heat exchange, and chemical and mechanical work performed in the phases of the cycle. \\
iii) We have verified the validity and self-consistency of previous derivation(s) of the efficiency at maximum power, by taking into account explicitly possible nonlinearities in the transport laws and possible density or chemical potential dependencies of the current in the transport equation. We have done this by proposing a transport model characterized by effective transport constants and an arbitrary skew-symmetric function of two variables, the chemical potentials of the two reservoirs. The precise form of this function is irrelevant, apart from the requirement that it produces a linear response to a small driving force. We conclude that  the efficiency at maximum power takes the value 1/2, predicted in earlier works based on selected models, \textcolor{black} {and establish the extent of its universality (see iv) and v) below)}.\\
iv)  We have discussed the correction terms, in the form of powers of the chemical potential difference, to this universal constant. We have found that the first-order correction vanishes for chemical engines with left/right symmetry or chemical engines featuring processes for which the skew-symmetric function depends only on the chemical potential difference.\\
v) We have tested the robustness of the value 1/2 for the efficiency at maximum power, by modifying the transport law to one that produces a purely nonlinear response. To this end, we have introduced a model featuring an intrinsically nonlinear transport equation, and have found that the efficiency at maximum power depends in a simple algebraic manner on the exponent of the nonlinearity.  Physical systems that can be described by a vanishing linear transport coefficient have not been identified in this paper. We leave this as a challenge for further research \cite{future}.\\
vi) We have analyzed our chemical engine from the point of view of weak dissipation. In this limit full consistency with entropy production notions is found and the result 1/2 for the efficiency at maximum power emerges naturally (for the generic transport law).

\section*{ACKNOWLEDGEMENTS}
We thank Kenichiro Koga for a discussion on transport processes, Christian Maes for his open mind towards purely nonlinear transport laws and Mehran Kardar for suggesting to include also sublinear transport models in our considerations. H.H. is supported by the FWO (Fonds voor Wetenschappelijk Onderzoek - Vlaanderen) and A.S. and J.O.I. by KU Leuven Research Grant OT/11/063.

\appendix
\section{The thermal cycle}
On the subject of the thermal cycle (without particle exchange) between temperatures $T_{high}$ and $T_{low}$ we can be concise. Firstly, it is readily checked that the Carnot efficiency $\eta_C = 1 - T_{low}/T_{high}$ is, as it should, the same for a Clausius gas and an ideal gas of point particles. It suffices to verify that, for the Clausius gas, isothermal processes are described by $p V_{\rm free} = constant$, and adiabatic processes by $p V_{\rm free}^{\gamma} = constant$ or $T V_{\rm free}^{\gamma -1} = constant$, with $\gamma = C_p/C_V$ the familiar specific heat ratio and $V_{\rm free} \equiv V-Nv_0$. Given this preliminary check, it is straightforward to see that the derivation of Curzon and Ahlborn of the efficiency at maximum power $\eta_{C,max}$ is not influenced in any way by a modification of the intrinsic characteristics of the working fluid. Indeed, the heat transfer coefficients in their derivation depend solely on the thickness $L$ and thermal conductivity $\kappa$ of the (solid) wall of the vessel enclosing the working fluid. Consequently, their main result 
 \begin{equation} \label{etaCA}
\eta_{C,max} = 1 - \sqrt{\frac{T_{low}}{T_{high}}} = \frac{1}{2}\eta_C + {\cal O}\left ((T_{high}-T_{low})^2\right )
\end{equation}
is independent of the precise nature of the working fluid.
Note that their main result is strictly speaking only valid to first order in $T_{high}-T_{low}$ because they assumed that heat fluxes through the vessel containing the working fluid are proportional to the temperature difference $\Delta T$ across the vessel wall (linear approximation). Specifically, they made the assumption
 \begin{equation} \label{F}
F = \alpha \Delta T,
\end{equation}
with $F$ the heat flux and $\alpha \approx \kappa / L $ the thermal conductivity per unit thickness of the wall. Note that $\kappa$ is defined through the more precise differential formulation,
 \begin{equation}
F = - \kappa \frac{dT}{dx}.
\end{equation}
Obviously, higher-order terms in $\Delta T$ are neglected in \eqref{F}, which can therefore only be expected to be reliable for small enough $\Delta T$. These higher-order terms generate higher-order terms in $T_{high}-T_{low}$ in the power and in the efficiency. In conclusion, \eqref{etaCA} is, strictly speaking, only reliable to first order in $T_{high}-T_{low}$. 


\begin{thebibliography}{100}
\bibitem{vanderWaals} \textcolor{black}{Our results for the efficiency at maximum power in this paper are also valid for a gas with van der Waals intermolecular attractions. Only the thermodynamic quantities in the steps of our cycle are modified. The net work performed over the entire cycle is unchanged. However, we wish to avoid the complication of possible phase transitions in our working fluid, and have opted for the simplest ideal-gas-like equation of state that explicitly contains the effects of a non-zero particle size.}
\bibitem{isocardinal} \textcolor{black}{This is a meaningful term because the ``cardinal number" is a number denoting quantity (but not order) in a group.}
\bibitem{valve} \textcolor{black}{Our reason for including an explicit dependence on the reservoir index $i$ in the effective transport coefficient $\kappa_i$ is that  the input valve (connecting the system to reservoir $a$) and the output valve (connecting it to reservoir $b$) may have different opening areas $A_i$ and opening lengths $L_i$.}
\end{thebibliography}
\end{document}